\documentclass[aps,twocolumn,showpacs,amssymb]{revtex4}

\usepackage[english]{babel}
\usepackage{amssymb}
\usepackage{indentfirst}
\usepackage{graphics}
\usepackage{amsmath}
\usepackage{amsthm}

\newcommand{\one}{\mbox{$1 \hspace{-1.0mm}  {\bf l}$}}
\def\tr{\mathrm{tr}}
\def\ket#1{\left| #1\right>}
\def\bra#1{\left< #1\right|}
\newcommand{\proj}[1]{\ket{#1}\bra{#1}}
\newcommand{\bea}{\begin{eqnarray}}
\newcommand{\eea}{\end{eqnarray}}

\usepackage{CatEn}

\usepackage{amsfonts,epsfig,color,bbm}

\begin{document}

\title{Quantum Kolmogorov Complexity and its Applications}

\author{Caterina E. Mora$^1$, Hans J. Briegel$^{1,2}$, and Barbara Kraus$^2$}

\affiliation{$^1$ Institut f\"ur Quantenoptik und Quanteninformation
  der \"Osterreichischen Akademie der Wissenschaften, Innsbruck,
  Austria\\
$^2$ Institut f{\"u}r Theoretische Physik, Universit{\"a}t Innsbruck,
Technikerstra{\ss}e 25, A-6020 Innsbruck, Austria}

\date{\today}

\begin{abstract}
Kolmogorov complexity is a measure of the information contained in a
binary string. We investigate here the notion of quantum
Kolmogorov complexity, a measure of the information required to
describe a quantum state. We show that for any definition of
quantum Kolmogorov complexity measuring the number of classical
bits required to describe a pure quantum state, there exists a
pure n-qubit state which requires exponentially many bits of
description. This is shown by relating the classical communication
complexity to the quantum Kolmogorov complexity. Furthermore we give 
some examples of how quantum Kolmogorov complexity can be applied to
prove results in different fields, such as quantum computation and
thermodynamics, and we generalize it to the case of mixed quantum states.
\end{abstract}
\maketitle

\section{Introduction}

In 1948, Shannon \cite{CEShannon} laid the foundation of modern
theory of information and communication by defining mathematically
the concept of information. {In particular, Shannon was interested
in the amount of information contained in typical messages emitted
from a source (a random variable). The Shannon entropy of a random
variable $X$ measures, in fact, how much information we gain on
average if we learn the value of $X$. A different, but
related issue, regards the idea of quantifying the information
contained in a single message (or, more generally, a single
object), independently of the statistical properties of the source
that emits it. This problem has been first considered by
Solomonoff in 1960, but its most relevant developments are due to
independent works by Kolmogorov and Chaitin \cite{kolmogorov,
chaitin}. The underlying idea, common to all these approaches, is
that the amount of information contained in a finite object (bit
string) is the size of the shortest program that, without
additional data, computes the string and halts. To give
full credit to all those who made pioneering contributions to the subject, one
should probably refer to such a measure of information as
``Solomonoff-Kolmogorov-Chaitin complexity''. For brevity we adopt the most common choice in literature, and refer
to it simply as Kolmogorov complexity.

The Kolmogorov complexity of a $n$-bit string,
$\bomega$, is defined as the length of the shortest binary program
that computes that string, i.e.

\beq K_U(\bomega)=\min_p\{l(p)|U(p)=\bomega\}
\label{ClassicalDefinition}, \eeq

where $U$ denotes a Turing machine, which computes $\bomega$,
given the program $p$ as input. One can prove that the Kolmogorov
complexity, also called algorithmic, or descriptive complexity, is
largely independent of the particular choice of the Turing machine. That is, changing Turing machine will
change the algorithmic complexity at most by a constant, which
becomes negligible in the limit of very long strings
($n\to\infty$) \cite{li}. For this reason one usually omits the
subscript ``$U$'' and writes simply $K(\bomega)$, or $K_\cl(\bomega)$ to underline the fact that we are considering the classical quantity.

Note that this quantity is not computable \cite{Thomas}. However,
it turned out that already upper bounds on the
Kolmogorov complexity of asymptotically long bit strings, where they can be computed, are very
fruitful. A trivial upper bound on the Kolmogorov complexity of
an $n$--bit string $\bomega= \omega_1,\ldots ,\omega_n$ is $n$,
since there always exists a program of the form: "Print
$\omega_1,\ldots ,\omega_n$". A string $\bomega$ is said to be
compressible by $k$ bits if there exists a program with less than
or equal to $n-k$ bits that gives $\bomega$ as an output. A simple
counting argument shows that at most $2^{k}$ strings have a
complexity smaller than or equal to $k$. Thus, for any $n$, there
exists at least one $n$-bit complex (incompressible) string.
Actually, one can show that almost all bit strings are complex
\cite{Thomas}.

The definition of Kolmogorov complexity satisfies the intuition
that it is much easier to describe a regular object than a random
one. Moreover, it is a good measure of the information content of
a bit string. A complex object contains a large amount of
information (albeit not necessarily structure \cite{bennettcomplexity}) 
and thus cannot be compressed. On the contrary,
regular strings have a much smaller complexity: for example, the
complexity of a periodic string is at most logarithmic in its
length. This logarithmic dependency is only due to the fact that
one has to specify the number $n$, which requires $\log(n)$ bits
of description. Notice that this dependency has no
deeper meaning than the mere specification of the length of the
output string. One can thus remove it by considering the conditional quantity
$K(\bomega | n)$, which measures the number of bits required to
describe $\bomega$ given the fact that the computer is given the
length $n$ of the string \footnote{Analogously, any definition of the quantum Kolmogorov complexity of an $n$-qubit state will have a $\log n$ term that specifies the number of qubits of the states and that can be removed by considering the conditional complexity.}.

The theory developed in the context of Kolmogorov complexity has
become very powerful in classical information theory. It has been
successfully employed as a general proof method, known as
``incompressibility method'' \cite{li}, in as diverse fields as
learning theory, complexity theory, and combinatorics, to name just
a few. The notion of Kolmogorov complexity has been used \cite{chaitin} to solve
in a simple and elegant way many otherwise difficult problems, such as, e.g.,
G\"odel's theorem \cite{goedel}. Recently it has also been used to
establish a connection between classical information and
thermodynamics \cite{BennettThermo, ZurekNature, Caves1, Caves2}. How to
generalize the concept of Kolmogorov complexity to the quantum
world has become a natural question with the development of the
theory of quantum information. Not only is a good notion of quantum
Kolmogorov complexity of deep fundamental interest, but
one also hopes it will allow to develop a theory similarly powerful as its
classical counterpart.

In recent years, a number of different definitions of quantum
Kolmogorov complexity have been proposed \cite{vitanyi, vandam,
gacs, prlnostro}, each generalizing the classical quantity in a
different way. Some measure  the number of qubits required to
describe a quantum state, while others count instead the number of
classical bits. The aim of this paper is to shed new light on the
different definitions of quantum Kolmogorov complexity. In
particular the results presented here will allow us to better
understand the physical meaning of some of the definitions. Furthermore we
will show how Kolmogorov complexity can, also in the quantum scenario, be
used as a proof method.

The paper is organized as follows. In Section II we review some of the
existing definitions of quantum Kolmogorov complexity of pure
states. Then we review the notion of communication complexity. We
will focus on the equality problem and, more precisely, on its
solution by means of the fingerprinting protocol studied in
\cite{fingerprinting}. In Section III we relate the number of
classical bits required to describe a quantum state (as a general
definition) to a problem in communication complexity. This allows us to
prove that there exist quantum states of $n$ qubits whose
description require exponentially many classical bits. In Section
IV we use the relation studied in Section III to derive some
implications for any definition of Kolmogorov complexity, including a natural generalization
to mixed states. Furthermore we will provide a non-trivial relation between the
so--called network Komogorov complexity of a class of pure quantum states
\cite{nostrolungo} and the classical Kolmogorov complexity of bit strings.
In the last section we apply the notion of
quantum Kolmogorov complexity to a number of problems, including
computation complexity and thermodynamics.

\section{Review of Quantum Kolmogorov Complexity and Communication Complexity}

In this section we first review the different definitions of
quantum Kolmogorov complexity and their properties. Second, we
review the notion of communication complexity. In particular we
will describe the fingerprinting protocol studied in
\cite{fingerprinting} to solve the equality problem. In Section
III we will relate these two concepts to each other. This section
will also serve to introduce our notation.

\subsection{Quantum Kolmogorov Complexity}

Many classical quantities, e.g the Shannon entropy, have been
successfully generalized to quantum information and have become
useful and powerful tools to understand and further develop quantum
information theory. In the case of Kolmogorov complexity, though,
the way to do so is not straightforward. Indeed, recently a number of
different definitions for quantum Kolmogorov complexity have been
introduced by G\`acs \cite{gacs}, Berthiaume et al. \cite{vandam},
Vitanyi \cite{vitanyi}, and Mora and Briegel \cite{nostrolungo}.
Contrary to the classical case, where different definitions have
been shown to be equivalent \cite{li}, in the quantum setting they
are not. The definition proposed in \cite{gacs} differs from the
others in that it is a generalization of an alternative definition
of algorithmic complexity, introduced by Levin \cite{levin}.
The general idea of such approach is not to search for the
shortest program that reproduces the string $\boldx$ (as one
does in the definition proposed by Kolmogorov), but rather to look
at the probability of obtaining $\boldx$ when a random program (that is, a sequence of 0 and 1
generated, e.g., by coin tosses) is given as input to a
Turing machine. It has been shown that the
negative logarithm of such a probability coincides with the
Kolmogorov complexity of the string $\boldx$. The probability
distribution over the set of binary strings generated by running
random programs on a Turing machine is called \emph{universal
probability}, and it can be characterized by precise mathematical
properties. In his work, G\`acs proposes a universal density
matrix, ${\boldsymbol{\mu}}$, that can be viewed as a
generalization of the universal probability distribution. The
definition of the Komogorov complexity of a quantum state
follows, thus, quite naturally from the classical theory. The two
possible choices, whose properties are analyzed in detail in
\cite{gacs}, are 
\beq
\overline{H}(\sphi)=-\langle\phi|\log\boldsymbol{\mu}\sphi{\textrm{~~and~~}}\underline{H}(\sphi)=-\log\langle\phi|\boldsymbol{\mu}\sphi~.
\eeq 
It is easy to see that, given an $n$-qubit state $\sphi$,
both quantities are upper bounded by $n$.

The other definitions mentioned above are, instead, based on a
generalization of complexity, as introduced by
Kolmogorov and Chaitin. In \cite{vandam} the Kolmogorov complexity
of a quantum state $\rho$ is defined as the size of the smallest
quantum program (state) that, given as input to a quantum Turing
machine, yields $\rho$ as output. In this work, the ``length of a
quantum state'' is defined as the logarithm of the dimension of
the smallest Hilbert space that contains the state itself. One can
then define the complexity of $\rho$ as 
\beq
K_{\textrm{BLV}}^{\epsilon}(\rho)=\min_{\pi}\{l(\pi) |
F(U(\pi),\rho)>{1-\epsilon} \}~, 
\eeq 
where $1-\epsilon$ measures
the fidelity, $F$, with which the state $\rho$ must be reproduced.
It is clear from the definition, that this quantity is a measure
of the \emph{quantum information} contained in a state. One can
prove that the complexity of an $n$-qubit state is upper bounded
by $n$.

In \cite{vitanyi}, on the contrary, the author proposes a
definition of quantum Kolmogorov complexity based on the
\emph{classical description} of the state. More precisely, the
complexity of a state $\sphi$ is defined as 
\beq
K_{\textrm{Vit}}(\sphi)=\min_p\{l(p) - \log|\langle
\psi|\phi\rangle|^2 | \mcu(p)=|\psi\rangle\}. 
\eeq 
The idea here
is to compute $\ket{\psi}$ which approximates the desired state
$\sphi$, and the term $- \log|\langle \psi|\phi\rangle|^2$
penalizes for a bad approximation of the state. Analogously to the
definition of Berthiaume et al., the complexity of a state $\sphi$
can grow at most linearly with the number of qubits whose state is
described by $\sphi$.

Another approach, which is related to Vitanyi's one has been
introduced by Zurek, first, and later by Caves and Schack
\cite{ZurekPRA, Caves1, Caves2, CavesSchack}. Since we will
discuss this approach in more detail in the last section of this
paper, we also want to briefly review it here. Suppose that on is given
a set of possible states, ${\cal S}=\{\ket{\Psi_i}\}$ and an
associated probability distribution $P=\{p_i\}_i$, with
$p_i=P(\ket{\Psi_i})$. In order to specify a particular element of
this set, one just has to specify the index, which is a purely
classical task. It is well known that the so--called Shannon-Fano
code can be used to obtain a (almost optimal) description
\cite{Thomas}. In order to explain the code, let us assume that
$p_1\geq p_2,\ldots$ and define $F_i=\sum_{k=1}^{i} p_k$. The codeword
$c_i$ for index $i$ is $r(F_i)$, which is $F_i$ rounded off to
$l_i=\lfloor\log_2(1/p_i)\rfloor$, i.e. the floor of
$\log_2(1/p_i)$, bits. Thus, if the information of the set ${\cal
S}$, also called the background information \cite{CavesSchack} is
given, then only $l_i$ bits are additionally required to describe
the state $\ket{\Psi_i}$ \footnote{Note that this is usually the
approach taken in classical theory, where the background
information of ${\mathcal{S}}$ is always small.}. Note that the
logarithmic term in the definition before, can be understood in
this way \cite{vitanyi}.

A completely different approach, based on the identification of a
state with its abstract preparation procedure, was investigated in
\cite{prlnostro, nostrolungo}. In this case the Kolmogorov
complexity of a state $\sphi$ is defined as the amount of the
classical information required to describe a circuit that prepares
$\sphi$ with the required precision $\epsilon$ from a fixed
initial state, e.g. $\ket{0}^{\otimes n}$. To be more precise,
first a complete and finite gate basis $B$ is fixed. Any state
$\sphi$ can then be prepared, with arbitrary finite precision,
$\epsilon$, by a circuit, $\mcc^{B,\epsilon}(\sphi)$, built with
gates from such a basis \footnote{As in \cite{nostrolungo}, here
and in the following we will say that circuit $\mcc^{B,\epsilon}$
prepares $\sphi$ with precision (or accuracy) $\epsilon$ if
$\langle\phi|\mcc^{B,\epsilon}(|0\rangle^{\otimes
n})|^2\geq1-\epsilon$}. Furthermore, as the basis is finite, it is
possible to define a finite code and to encode any circuit in a
classical sequence. Thus it is possible to associate to each state
$\sphi$ a set of strings $\bomega^{B,\epsilon}(\sphi)$, each of
which encodes a circuit that prepares $\sphi$ with the required
accuracy. The Kolmogorov complexity of $\sphi$ is then defined as
the minimum among the classical Kolmogorov complexities of the
strings $\bomega(\mcc^{B,\epsilon})(\sphi)$ \beq
K_\net^{B,\epsilon}(\sphi)=\min_{\mcc^{B,\epsilon}(\sphi)}
K(\bomega(\mcc^{B,\epsilon}(\sphi))). \label{DefNetwork} \eeq
Here, the minimum is taken over all circuits, $\mcc^{B,\epsilon}$,
such that $|\langle\phi|\mcc^{B,\epsilon}(|0\rangle^{\otimes
n})|^2\geq 1-\epsilon$.  Taking into account that one needs at
most $2^n\log\frac{1}{\epsilon}$ gates in order to prepare an
$n$-qubit state, and that the length of
$\bomega^{B,\epsilon}(\sphi)$ is upper bounded by the number of
gates, one obtains that $K_\net^{B,\epsilon}(\sphi)$ is upper
bounded by $2^n\log\frac{1}{\epsilon}$, up to terms of smaller
order. From now on we will refer to this definition as
``network'' complexity.

Assuming that we restrict the choice of the set of basis gates to
1- and 2-qubit gates (differing from the original definition
presented in \cite{nostrolungo}), one can show that the network
complexity is independent (up to a constant) of the actual choice
of basis. Keeping this in mind, we will thus simply write
$K_\net^{\epsilon}(\sphi)$.

It should be noted here that the network complexity has not only a
clear physical meaning, since it gives a measure of how hard it is
to describe a way to prepare the state, but furthermore it was
also shown that there exists a connection between this definition
of Kolmogorov complexity and entanglement \cite{prlnostro}. In
order for a state to have maximal complexity (that is, exponential
in the number of qubits) it is in fact necessary that it is highly
entangled: more precisely, a complex state must have maximum
Schmidt measure \cite{hans}. On the contrary, a product state has
complexity at most linear in the number of qubits.

Another possibility to describe a pure state classically is simply
to describe the coefficients of the state in a certain basis, e.g.
the computational basis. We will refer to this complexity as 
computational-basis-expansion complexity, or, more briefely, 
CBE complexity. The CBE complexity of a $n$-qubit state
$\ket{\Psi}=\sum_{i_1,\ldots,i_n}c_{i_1,\ldots,i_n}\ket{i_1,\ldots,i_n}$,
with $i_j\in \{0,1\}$ and complex coefficients
$c_{i_1,\ldots,i_n}$ is then
\bea
K^\epsilon_{\textrm{CBE}}(\ket{\Psi})=K_\cl[(c_{0,\ldots 0},\ldots,c_{1,\ldots 1})].
\eea
In general, the coefficients $c_{i_1,\ldots,i_n}$ are computed 
only with a finite precision $\epsilon$, i.e. each of them is 
approximated by a number $\tilde{c}_{i_1,\ldots,i_n}$ such
that $\vert c_{i_1,\ldots,i_n}-\tilde{c}_{i_1,\ldots,i_n}\vert
\leq \epsilon$. This implies that the state that has been
described in this way, i.e.
$|\tilde{\Psi}\rangle=\sum_{i_1,\ldots,i_n}
\tilde{c}_{i_1,\ldots,i_n}\ket{i_1,\ldots,i_n}$, fulfills the inequality
$\Vert |\tilde{\Psi}\rangle- |\Psi\rangle\Vert^2\leq 2^n
\epsilon^2$. Thus, $1-|\langle\Psi\tilde{\Psi}\rangle|^2 \leq
2^{n-1}\epsilon^2$, which implies that the precision $\epsilon$  must scale like $1/2^{n/2}$,
similarly as the precision with which one must describe each gate
occurring in the definition of the network complexity.

Yet another possibility is to use a measurement and describe the state
by the outcome probability of this measurement. One might for
instance choose an ``algorithmically simple'' measurement, i.e. a
 POVM that can be described with only a few bits, and describe the 
 state by the outcome probability of this measurement.

As we have already mentioned, in classical information theory
the different definitions of complexity are provably
equivalent, while this does not hold in quantum theory. Indeed some of
the definitions described here have qualitatively different
behaviors, e.g. the scaling with $n$. In this paper we aim to shed new
light on the notion of quantum Kolmogorov complexity and to
better understand its physical meaning.

One of the main drawbacks of Kolmogorov complexity is the fact that it is not computable. Despite of this, though, it is possible to give upper bounds for it, and these are sufficient and instrumental for many applications. Furthermore, we recall that even in the case of computable quantities, such as, e.g., communication or computation complexities,  one is often not interested in the exact value, since it might be very hard to compute. In these cases too, one
is mainly concerned with the scaling of these quantities as a
function of the input size, $n$. Owing to this fact, the following
notation is widely used. One says that a function, $g(n)$, is
${\cal O}(f(n))$ if there exist two constants, $c\in{\mathbb{R}}$ and $n_0\in{\mathbb{N}}$ such
that $g(n)\leq cf(n)$ for all $n\geq n_0$. That is, for
sufficiently large $n$, $g$ is upper bounded by $f$. Similarly one
says that a function $g(n)$ is $\Omega (f(n))$ if there exist
two constants, $c$ and $n_0$ such that $g(n)\geq cf(n)$ for all
$n\geq n_0$. If the behavior of two functions, $g,f$ is
asymptotically the same, up to a constant factor, i.e. if $g$ is both
${\cal O}(f(n))$ and $\Omega (f(n))$ then we say that $g$ is
$\Theta (f(n))$.

From now on, we will call a bit string (a state)
algorithmically simple when its Kolmogorov complexity is at most ${\cal O}(\log(n))$ (at
most $\mco(n)$).


\subsection{Communication complexity}
\label{SectionCommCompl}

The theory of communication complexity deals with the following
type of problem. Two distant parties, Alice and Bob, hold each an
input, $\boldx$ and $\boldy$ respectively (usually one has
$\boldx,\boldy\in\{0,1\}^n$). The common goal of the parties is to
compute the value $f(\boldx,\boldy)$ for a given function $f$
known to both parties. The \emph{communication complexity} of a
function $f$, $C_{Cl}(f)$, is defined as the minimum number of
bits that Alice and Bob need to communicate in order to evaluate
$f(\boldx,\boldy)$.

In the quantum setting, the communication between Alice and Bob
takes place through a quantum channel. Analogous to the
classical case, the quantum communication complexity of $f$,
$C_{Q}(f)$, is the minimum number of qubits that Alice and Bob
need to communicate to evaluate $f(\boldx,\boldy)$.

A slightly different communication model is the so--called \emph{simultaneous
message passing} model, first introduced by Yao \cite{yao_sim}. In
this model there is a third party --the \emph{referee}-- in
addition to Alice and Bob. As before, the common aim of the
parties is to compute $f(\boldx,\boldy)$, where $f$ is a given function and
$\boldx$ and $\boldy$ are the input strings held by Alice and Bob
respectively. In this case, however, no direct communication between Alice
and Bob is allowed. They can only send messages to the referee,
who will then compute $f(\boldx,\boldy)$ based on the information he has
received.

There are many cases in which one does not require the function
$f$ to be computed with certainty, but allows a certain failure
probability, $\delta$. The conditions on this error, as well as
other factors, distinguish different scenarios in which
communication complexity problems are usually solved (see e.g.
\cite{dewolf} for an overview). We will focus here on the
so--called \emph{worst case scenario}. In this case, one is
interested in the amount of communication needed to compute
$f(\boldx,\boldy)$ correctly with probability larger or equal to
$1-\delta$, for any $\boldx$ and $\boldy$.

Of course, any function $f$ can be exactly evaluated if both Alice
and Bob send their full input to the referee, who can then compute
$f(\boldx,\boldy)$. It follows thus that $2n$ is an upper bound
for any communication problem in the simultaneous message passing
model \footnote{Note that the same argument holds when direct
communication between Alice and Bob is allowed. In this case,
though, the bound is $n$, and not $2n$, since only one of the
parties need to communicate his/her input}. To improve such a
protocol, one might consider the case in which Alice and Bob send
to the referee a \emph{compressed} version of their input. In this
case, the communication required depends on the Kolmogorov
complexity of the two inputs. If one considers the worst--case
scenario and does not allow any error, this does not help to
improve the upper bound. By considering the case in which
the inputs are incompressible $n$--bit strings, in fact, it is easy
to see that $2n$ bits of communication are still needed
\footnote{Such a compression, though, might be useful if one is
interested in other scenarios, or if the domain of $\boldx$ and
$\boldy$ is restricted in some way.}. However, if a certain
failure probability is tolerated, one can show that only ${\cal
O}(\sqrt{n})$ bits of communication are required (see below).

Even though Holevo's Theorem states that by sending $m$ qubits one
cannot convey more than $m$ classical bits of information, it has
been shown that there are cases in which the quantum setting is
significantly more powerful than the classical one (see e.g.
\cite{dewolf} for an overview). In fact, there are functions for
which the classical communication complexity is exponentially
larger than the quantum one. One such instance, which we will
consider in the following, is the equality problem. This is a
worst--case simultaneous message passing -- model where the
function to be evaluated is \beq
\textit{EQ}_n(\boldx,\boldy)=\left\{\begin{array}{c c} 1,~&\textrm{if}~\boldx=\boldy\\
0,~&\textrm{if}~\boldx\neq \boldy~,
\end{array}\right.
\eeq
with $\boldx,\boldy\in\{0,1\}^n$.\\

It has been proved \cite{ambainis, newman} that, allowing
a small probability of failure, only ${\cal O}(\sqrt{n})$ bits of classical communication are
needed to compute $\textit{EQ}_n(\boldx,\boldy)$. Furthermore, this amount of communication has been shown to be both
necessary and sufficient for the solution of the equality problem.

Let us now briefly recall one optimal classical protocol here. It
is known that for every $n\in {\mathbb{N}}$ and for a fixed $c>1$, and
$0<\Delta<1$, there exists an error correcting code $E:\{0,1\}^n
\to \{0,1\}^m$, with $m=cn$, such that the distance between code
words ${\mathbf{E(x)}}$ and ${\mathbf{E(y)}}$ is at least
$(1-\Delta)m$ \footnote{An example, for instance, is given by
Justesen codes. In this case one may choose any $c>2$ and
$\Delta<9/10+1/(15c)$}. Alice and Bob apply a given error
correcting code (where the specification of the code is
independent of $n$) to $\boldx$, $\boldy$ respectively. Then they
choose randomly an index $i$, $j$ $\in \{1,\ldots,m\}$,
respectively. Alice (Bob) sends $i$ and the $i$--th bit value of
${\mathbf{E(x)}}$, i.e. $E_i(\boldx)$, ($j$, and $E_j(\boldy)$) to
the referee. If $i\not=j$ the referee cannot conclude anything and
tells Alice and Bob to restart the protocol. However, if $i=j$,
the referee checks whether $E_i(\boldx)$ equals $E_j(\boldy)$. If this
is the case he concludes that $\boldx=\boldy$ else he concludes
$\boldx\not=\boldy$. The probability of having a collision, i.e.
that the random indices $i,j$ coincide, is $1/m$. The probability
of making an error is $P_{error}= P(E_i(\boldx)=E_i(\boldy)\mid
\boldx\not=\boldy)/m<\Delta$. Thus, in order to make the error
probability independent of $m$ (i.e. independent of $n$), Alice
and Bob have to send ${\cal O}(\sqrt{n})$ bits. Note that the
error can be reduced to any $\delta >0$ if Alice and Bob send
${\cal O}(\log(1/\delta))$ random bits of the code words, which
implies that the communication complexity of the equality problem,
with error probability $\delta$, is ${\cal
O}(\sqrt{n}\log(1/\delta))$.

In \cite{fingerprinting}, the case in which Alice and Bob's
fingerprints can consist of \emph{quantum} information is studied
and it is shown that $\mco(\log_2 n)$-qubit fingerprints are
sufficient to solve the equality problem (with arbitrarily small
error probability). Let us briefly review also this protocol here.
Using the same notation as before one defines, for each
$\boldx\in\{0,1\}^n$ the following $(\log_2 m +1)$-qubit state
\beq |h_\boldx\rangle=\frac{1}{\sqrt{m}}\sum_{i=1}^m |i\rangle
|E_i(\boldx)\rangle~, \eeq where, as above, $E_i(\boldx)$ is the
$i$-th bit of the word ${\boldE(\boldx)}$. Since the two words,
${\boldE(\boldx)}$ and ${\boldE(\boldy)}$ can be equal for at most
$m \Delta$ positions, for any $\boldx\neq \boldy$ we have
$\vert\langle h_\boldx|h_\boldy\rangle\vert \leq \Delta$. The
simultaneous message passing protocol for the equality problem is
the following.
\begin{enumerate}
\item Alice and Bob, holding $\boldx$ and $\boldy$ respectively,
prepare a quantum system in the states $|h_\boldx\rangle$ and
$|h_\boldy\rangle$ respectively and send it to the referee. \item
The referee uses an auxiliary qubit, $R$ and prepares the state
$(H_R\otimes{\mathbb{I}}_{AB})({\textrm{C-SWAP}}_{R,AB})
(H_R\otimes{\mathbb{I}}_{AB})|0\rangle_R(|h_\boldx\rangle|h_\boldy\rangle)_{AB}=
1/2[\ket{0}_R(\ket{h_\boldx,h_\boldy}+\ket{h_\boldy,h_\boldx})_{AB}+\ket{1}_R(\ket{h_\boldx,h_\boldy}-\ket{h_\boldy,h_\boldx})_{AB}]$.
Then he measures the auxiliary system in the computational basis
and associates $\boldx=\boldy$ with the outcome ``$0$'', and
$\boldx\not=\boldy$ with the outcome ``$1$''.
\end{enumerate}
With this test, the referee can determine the value of
$\textit{EQ}_n(\boldx,\boldy)$ with error probability
$(1+\Delta^2)/2$. Such a probability can be reduced to any
$\delta>0$ by considering the fingerprint
$|h_\boldx\rangle^{\otimes k}$, with
$k=\mco(\log_2\frac{1}{\delta})$. In this case the length of each
fingerprint is $\mco(\log_2 n \log_21/\delta)$.

\section{Relation between classical communication complexity and quantum
Kolmogorov complexity}

In this section we establish a relation between Kolmogorov
complexity and communication complexity. To this end, we consider
a classical protocol that simulates the quantum fingerprinting
protocol. That is, instead of sending quantum systems (prepared in
the states $\ket{h_\boldx}, \ket{h_\boldy}$), Alice and Bob send
to the referee a classical description of these states. Since
there exists a lower bound on the classical communication
complexity for this problem, this consideration leads to a lower
bound to the number of bits required to describe those states,
i.e. to their quantum Kolmogorov complexity. We show now that this
lower bound is exponential in the number of qubits.

Let $K_Q^\epsilon$ be an arbitrary definition of the quantum
Kolmogorov complexity with the following physical meaning: given
$K_Q^\epsilon(\sphi)$ classical bits one can prepare the state
$\ket{\psi}$ with precision $\epsilon$. Now and in the
following, we say that $K_Q^\epsilon(\sphi)$ bits of information allow to prepare
$\sphi$ with precision $\epsilon$ if it contains all the
information that is needed to prepare a state $\epsilon$-close to
$\sphi$ with a perfect apparatus. Or, equivalently, given
$K_Q^\epsilon(\sphi)$ bits of information, we can write down a
state, $\epsilon$-close to $\sphi$. At the same time we imply that
such a task is impossible with a smaller amount of
information. We show now that for any such definition there
exists a $n$--qubit state $\ket{\psi}$ such that
$K_Q^\epsilon(\ket{\psi})$ is exponential in $n$, ruling out some
of the definitions reviewed in Section II.

Let $K_Q^\epsilon(|h_\boldx\rangle)$ and
$K_Q^\epsilon(|h_\boldy\rangle)$ be the Kolmogorov complexities
of the two fingerprints used in the protocol. Then a possible
classical communication protocol is the following:
\begin{enumerate}
\item Given $\boldx$ and $\boldy$, respectively, Alice and Bob
send the description of the $(\log_2 m +1)$-qubit states
$|h_\boldx\rangle$ and $|h_\boldy\rangle$ to the referee,
requiring
$K_Q^\epsilon(|h_\boldx\rangle)+K_Q^\epsilon(|h_\boldy\rangle)$
bits of communication.
\item In order to distinguish between the
case where $|h_\boldx\rangle=|h_\boldy\rangle$ or
$|h_\boldx\rangle\neq|h_\boldy\rangle$, the referee could e.g. simulate (on
a classical computer) the circuit described in the quantum
fingerprinting protocol.
\end{enumerate}
The precision with which the referee is able to simulate the
quantum circuit, and thus the probability with which he can obtain
the correct answer, depends on the precision with which the two
quantum states are described. It is straightforward to show that a
final error probability (such as in the quantum fingerprinting
protocol) $\Delta$, is obtained if the states are described with
precision $\epsilon\leq\sqrt{\Delta}$ \footnote{This is clear as
all the states coded by the same word are almost parallel, while
those who appear in the error correcting code are almost
orthogonal.}.

The number of bits of communication of this protocol is
$K_Q^\epsilon(|h_\boldx\rangle)+K_Q^\epsilon(|h_\boldy\rangle)$.
As we know that the optimal bound for the classical communication
complexity of the equality problem, in the simultaneous passing
model, is $\sqrt{n}$, it follows that there must be at least a
case for which
$K_Q^\epsilon(|h_\boldx\rangle)+K_Q^\epsilon(|h_\boldy\rangle)
\geq \mco(\sqrt{n})$. Considering that $m=nc$, it follows that
there must exist an $M$-qubit state
$|\Psi_M^{\textit{Compl}}\rangle$, with $M=\log_2 m$ such that
$K_Q^\epsilon(|\Psi_M^{\textit{Compl}}\rangle)\geq
\mco(2^{M/2}/c)$. This shows that any definition of the quantum
Kolmogorov complexity that counts the number of classical bits 
required to prepare a state must grow exponentially with the number of qubits.

One of the definitions presented in Sec II that fulfills the
necessary requirement of growing exponentially with the number of
qubits is the network complexity \cite{prlnostro}. Note that also
the CBE complexity fulfills this condition.

\section{Implications for Classical and quantum Kolmogorov complexity}

The considerations of the previous section allow us to derive some
implications regarding the Kolmogorov complexity. The result, that the
 complexity of a pure $n$--qubit state can be $2^n$,
holds for any definition of the quantum Kolmogorov complexity that
measures the amount of classical bits required to prepare the
state. Since a state can also be described by a
outcome--probability distribution of local measurements, this
bound leads also to a bound on a classical Kolmogorov complexity
of a probability distribution. This consideration then guides us
how to define the quantum Kolmogorov complexity for mixed state.
Furthermore, we show that the Kolmogorov complexity of the states
$\ket{h_\boldx}$ is essentially equal to the classical Kolmogorov
complexity of the classical bit string $\boldx$.

\subsection{Classical Kolmogorov complexity of a probability distribution and quantum Kolmogorov complexity of mixed states}

One possible description of a state is a complete measurement and
the probability distribution of the outcomes of this measurement.
Let us choose the measurement itself to be algorithmically simple,
e.g. a local measurement. Thus, the POVM we consider consists of
at most $2^{2n}$ elements, which can be easily described (the
length of the description depends at most logarithmically on $n$).
Since we consider a complete measurement, the complexity of the
quantum state is upper bounded by the classical Kolmogorov
complexity of its outcome probabilities, $P_{\Psi}$, i.e
$K_Q^\epsilon(\ket{\Psi})\leq K_{cl}(P_{\Psi})$, for
$\epsilon\rightarrow 0$. Note that the length of $P_{\Psi}$ is
upper bounded by $2^{2n}$. As we know that there exists a complex
pure state, $\ket{\Psi}$, this implies that there exists a complex
probability distribution $P_{\Psi}$, with $\Omega(2^n) \leq
K_{cl}(P_{\Psi}) \leq {\cal O}(4^n)$. Furthermore, any probability
distribution describing a complex state fulfills these
inequalities.

We now find some constraints on the quantum Kolmogorov complexity
of mixed states. Let us consider the pure state $\ket{\Psi}=\sum_i
\sqrt{p_i}\ket{i,i}$, with $p_i\geq 0$ and $\sum_i p_i=1$ and
where $\ket{i}=\ket{i_1,\ldots i_n}$ with $i_j\in \{0,1\}$ denotes
the computational basis. This state is a so-called purification of
the mixed state $\rho= \sum_i p_i \ket{i}\bra{i}$. $\rho$ can be
easily computed from $\ket{\Psi}$ and vice versa, requiring only
the additional information: "$\ket{\Psi}$ is a purification of
$\rho$, where the auxiliary system comprises the second set of
qubits. Furthermore the Schmidt-basis is the same for both
subsystems." Thus, the quantum Kolmogorov complexity of these two
states must be the same. One possible description of $\rho$ (or
$\ket{\Psi}$) is a description of the vector
$P=(p_1,\ldots,p_{2^n})$, containing all the eigenvalues of $\rho$
plus the information that $\rho$ is diagonal in the computational
basis. This implies the following relation between the Kolmogorov
complexities: \bea K^\epsilon_Q(\ket{\Psi})=K^\epsilon_Q(\rho)\leq
K_{cl}(P), \mbox{ for }\epsilon\rightarrow 0\eea where
$K^\epsilon_Q$ is any definition of the quantum Kolmogorov
complexity which allows us to describe the states in such a way
that they can be prepared.

Obviously, an arbitrary purification of a mixed state $\rho$, plus
the additional information of which systems are auxiliary systems
allows one to describe $\rho$. Therefore, for a general definition
of the Kolmogorov complexity it must hold that \bea
K_Q^\epsilon(\rho) \leq
\min_{\ket{\psi}}\{K_Q^\epsilon(|\psi\rangle_{AR}) |
\textrm{Tr}_R(|\psi\rangle_{AR}\langle\psi|)=\rho\}. \eea On the
other hand, it is clear that there exists at least one
purification of the state $\rho$, whose complexity is the same as
the one of $\rho$, namely the state whose Schmidt-basis is the
same for both subsystems and equals the eigenbasis of $\rho$.
These considerations strongly suggest that the generalization of
the Kolmogorov complexity to mixed states should be \bea
K_Q^\epsilon(\rho):=\min\{K_Q^\epsilon(|\psi\rangle_{AR}) |
\textrm{Tr}_R(|\psi\rangle_{AR}\langle\psi|)=\rho\}. \eea

Let us show now that the precision with which the mixed state is
described is exactly the same as the one with which its
purification is described. If $K_Q^\epsilon(|\psi\rangle_{AR})$
bits describe the state $|\tilde{\psi}\rangle_{AR}$, then $\vert
\langle\tilde{\psi}\vert\psi\rangle\vert ^2\geq 1-\epsilon$. Let
us now use Uhlmann´s theorem, which states that
$F(\rho,\sigma)=\tr[(\sqrt{\rho}\sigma\sqrt{\rho})^{1/2}]=\mbox{max}_{\ket{\Psi},\ket{\phi}}
\vert \bra{\Psi}\phi\rangle\vert$, where the maximization is over
all purifications $\ket{\Psi}$ of $\rho$ and $\ket{\phi}$ of
$\sigma$ \cite{nielsenchuang}. Therefore the reduced states
$\tilde{\rho}=\textrm{Tr}_R(|\tilde{\psi}\rangle_{AR}\langle\tilde{\psi}|)$
and $\rho=\textrm{Tr}_R(|\psi\rangle_{AR}\langle\psi|)$ fulfill
the inequality $F(\rho,\tilde{\rho})^2\geq \vert
\bra{\Psi}\tilde{\phi}\rangle\vert ^2=1-\epsilon$. Thus, if the
description of the purification of the mixed state leads to a
state which is $\epsilon$ close to the actual purification, then
the corresponding reduced states are also $\epsilon$ close.

As an example we can consider the (very simple) case of the
maximally mixed state of $n$ qubits ${\mathbb{I}}_A$. We will
choose the network complexity here. In this case it is easy to see
that its complexity is at most logarithmic in the number of
qubits. The $2n$-qubit state
$(|00\rangle+|11\rangle)_{AR}^{\otimes n}$ is a purification of
the maximally mixed state on $A$. To prepare such a state it is
sufficient to divide the $2n$ qubits into couples, apply to the
first qubit of each couple a Hadamard gate and then apply a C-NOT
gate between the two. The description of the circuit is extremely
easy, and depends on $n$ only in the specification of the total
number of qubits required. We have thus \best
K_\net^\epsilon(\frac{\mathbb{I}}{n})\leq \log
n+2\log\frac{1}{\epsilon}. \eest

The term in $\epsilon$ is separated from that in $n$ as there are
only two different types of gates required.


\subsection{Quantum complex states from algorithmically complex strings}

From the analysis of the fingerprinting protocol, we have seen
that there must exist a complex (and therefore highly entangled)
state of the form $|h_\boldx\rangle=\sum_{i=1}^m
|i\rangle|E_i(\boldx)\rangle$, where $\boldx\in\{0,1\}^n$ and
$E_i(\boldx)$ is the $i$-th bit of the encoding
$\boldE(\boldx)\in\{0,1\}^m$ of $\boldx\in\{0,1\}^n$ using code
$E$ (of the type described in section \ref{SectionCommCompl}).

While it is not necessarily true in general, there are cases in
which the codes $E$ have the additional property of being
algorithmically simple, i.e. such that $K_\cl(E)=\mco(\log(n))$.
This is true, e.g., for Justesen codes \cite{justesen}. When such
additional condition is satisfied, it is possible to prove an
important relationship between the classical Kolmogorov
complexities of $\boldx$, $\boldE(\boldx)$, and the quantum
(network) complexity of the states $|h_\boldx\rangle$. More
precisely, one finds that
$\displaystyle{K_\cl(\boldx)=\Theta(K_\cl(\boldE(\boldx)))=\Theta(K_\net^{B,\epsilon}(|h_\boldx\rangle))}$.
The proof of this statement consists of two parts. In the first
part (Observation \ref{lemmaxEx}) we show that the complexity of
the encoded word $\boldE(\boldx)$ coincides (up to a constant)
with that of $\boldx$. In the second part (Proposition
\ref{prophxEx}) we prove that the network complexity of the
quantum state $|h_\boldx\rangle$ is essentially the same as the
classical Kolmogorov complexity of $\boldE(\boldx)$.

\begin{obs} \label{lemmaxEx} If $E$ is an algorithmically simple code,
and $\boldE(\boldx)\in\{0,1\}^m$ is the encoding of
$\boldx\in\{0,1\}^n$, then
$K_\cl(\boldE(\boldx))=\Theta(K_\cl(\boldx))$
\end{obs}
\begin{proof}
To prove the statement, we show that $K_\cl(\boldE(\boldx))=\mco(K_\cl(\boldx))$ and $K_\cl(\boldx)=\mco(K_\cl(\boldE(\boldx)))$. \\
The first of the two equalities follows immediately by the fact that the code $E$ is algorithmically simple. To describe $\boldE(\boldx)$, in fact, it is sufficient to give the program for $E$, and a description of $\boldx$. Thus, $K_\cl(\boldE(\boldx))\leq K_\cl(E)+K_\cl(\boldx)=\mco(K_\cl(\boldx))$.\\
In order to prove the reverse statement, we consider that $\boldx$
can be described as the string whose encoding is $\boldE(\boldx)$.
Thus it is possible to prepare $\boldx$ by allowing a computer to
generate all possible strings $\boldE(\boldy)$ (for
$\boldy\in\{0,1\}^n$) and compare each of them with
$\boldE(\boldx)$, halting when $\boldE(\boldy)=\boldE(\boldx)$,
i.e., when $\boldy=\boldx$. This implies that $K_\cl(\boldx)\leq
K_\cl(\boldE(\boldx))+K_\cl(E)=\mco(K_\cl(\boldE(\boldx)))$.
\end{proof}

\begin{prop}
\label{prophxEx} Let $E$ be an algorithmically simple
error-correcting code, such that
$d(\boldE(\boldx),\boldE(\boldy))\geq(1-\Delta)m$, where
$\boldE(\boldx),\boldE(\boldy)\in\{0,1\}^m$ are the encodings of
$\boldx\in\{0,1\}^n$, $\boldy\in\{0,1\}^n$ respectively. Then, the
$(\log m+1)$-qubit state $|h_\boldx\rangle=\sum_{i=1}^m
|i\rangle|E_i(\boldx)\rangle$, where $\langle
i|j\rangle=\delta_{ij}$, has complexity
$\displaystyle{K^{\epsilon}_\net(|h_\boldx\rangle)=\Theta(K_\cl(\boldE(\boldx)))}$
for sufficiently small $\epsilon$.
\end{prop}
\begin{proof}
In order to prove the statement we will show that
$\displaystyle{K^{\epsilon}_\net(|h_\boldx\rangle)=\mco(K_\cl(\boldE(\boldx)))}$
and
$K_\cl(\boldE(\boldx))=\mco(K^{\epsilon}_\net(|h_\boldx\rangle))$.
Let us first prove that
$\displaystyle{K^{\epsilon}_\net(|h_\boldx\rangle)=\mco(K_\cl(\boldE(\boldx)))}$.
Consider the following circuit, that acts on the initial state
$|0\rangle^{\otimes(\log m +1)}$. Initially a Hadamard gate is applied
to each of the first $\log m$ qubits, generating the state
$((|0\rangle+|1\rangle)/2)^{\otimes \log m}\otimes |0\rangle$.
Then, one applies the unitary $E_i$ (such that
$E_i|0\rangle=|E_i(\boldx)\rangle$) to the last qubit, conditional
on the state of the first qubits being $|i\rangle$, i.e. the total operation being $|i\rangle|0\rangle \mapsto |i\rangle|E_i(\boldx)\rangle$. This operation
can be performed by a circuit composed of single-qubit and CNOT
gates, whose description requires at most $\mco((\log m)^2)$ bits
\footnote{This second part of the circuit can be described as
follows. For $i$ that goes from 0 to $m-1$, repeat the following
instructions. First apply a NOT gate to all the qubits whose
position corresponds to a zero in the binary description of $i$.
Second, apply the controlled-unitary $C^{\log m}(E_i)$ (as defined
in \cite{nielsenchuang}): this controlled unitary has a very
simple description (see, e.g., \cite{nielsenchuang}). Finally
apply NOT gates to all the qubits that were modified in the first
step (and reset all work qubits to zero)}. The state generated by
this circuit is $|h_\boldx\rangle$. The complexity of the state
$|h_\boldx\rangle$ is of course bounded by the complexity of the
circuit just described.
\beq
\begin{split}
K^{\epsilon}_\net(|h_\boldx\rangle) &\leq 2\log\frac{1}{\epsilon}
+ (\log m)^2 \log\frac{1}{\epsilon} + K_\cl(\boldE(\boldx)) \\
&=\mco(K_\cl(\boldE(\boldx))),
\end{split}
\eeq
when regarded as a function of $m$.
The first term takes into account the possibility that the
Hadamard and the {\textit{NOT}} gates might not be included in the
gate basis $B$. Note that this contribution is constant in $m$.
The second term takes into account the description of the
subcircuit in which each gate $E_i$ is applied conditional on the
state of the first $\log m$ qubits. The term
$K_\cl(\boldE(\boldx))$ is the one that specifies when
$E_i={\mathbb{I}}$ and when, instead, $E_i={\textit{NOT}}$. Note
that this proof remains valid even if the code $E$ is not
algorithmically simple. In the following we see that this property of the code
is needed in order to prove the reverse statement.

We show now that $K_\cl(\boldE(\boldx))=\mco(K^{\epsilon}_\net(|h_\boldx\rangle))$\\
Let $\bomega_h$ be the classical description of the circuit that
prepares $|h_\boldx\rangle$, such that
$K_\cl(\bomega_h)=K^{\epsilon}_\net(|h_\boldx\rangle)$, and
consider the following program.
\begin{enumerate}
\item Read $\bomega_h$. \item Simulate the circuit and compute the
approximate state $\vert \tilde{h}_\boldx\rangle$. \item For $i$
from 1 to $m$, compute $\langle i\vert \tilde{h}_\boldx\rangle$ .
\end{enumerate}

This is a purely classical program that, simulating the (exact)
creation of the quantum state $|h_\boldx\rangle$ and a number of
operations on it, allows to prepare the classical bit string
$\boldE(\boldx)$. The length of this program can be estimated as
follows. The first step requires
$\mco(K_\cl(\bomega_h))+\log K_\cl(\bomega_h)$ bits; the second
contributes a quantity that is constant in $m$ and depends
only on the precision with which we require the simulation to be
effectuated. The last step gives a contribution that is $\mco(\log
m)$. As this will not in general be an optimal program, its length
is an upper bound to the complexity of $\boldE(\boldx)$. \beq
\begin{split}
K_\cl(\boldE(\boldx))&\leq  \mco(K_\cl(\bomega_h))+\mco(\log n) + \mco(1)
=\\
&=\mco(K_\cl(\bomega_h))=\mco(K^{\epsilon}_\net(|h_\boldx\rangle)).
\end{split}
\eeq

In the analysis of the program given above, we have not yet taken
into account the effect of the error $\epsilon$ in the preparation
of the quantum state $|h_\boldx\rangle$. We do so now, assuming
that the circuit described by $\bomega_h$ does not prepare
$|h_\boldx\rangle$, but rather some state $\vert
\tilde{h}_\boldx\rangle$ such that $|\langle h_\boldx\vert
\tilde{h}_\boldx\rangle|^2\geq 1-\epsilon$. Without loss of
generality we write $\vert
\tilde{h}_\boldx\rangle=(1/\sqrt{m})\sum_{i=1}^m\alpha_{ij}|j\rangle|\tilde
E_i(\boldx)\rangle$. As a consequence, the program described above
will yield as final output a string $\tilde\boldE(\boldx)$, with
$\langle\boldE(\boldx),\tilde\boldE(\boldx)\rangle\geq m
\sqrt{1-\epsilon}$, where
$\langle\boldE(\boldx),\tilde\boldE(\boldx)\rangle$ is the scalar
product between the two $m$-dimensional vectors whose coefficients
(in an orthonormal basis) are given by $\boldE(\boldx)$ and
$\tilde\boldE(\boldx)$ respectively.

Since $E$ is chosen to be algorithmically simple, it is possible
to verify whether or not the final string $\tilde\boldE(\boldx)$
is part of the code. If it is not so, then it follows immediately
that there has been an error somewhere, which, if $\epsilon$ is
sufficiently small (depending on the error correcting code) can be
corrected. The only case in which the program described above
cannot correct itself is when the output word
$\tilde\boldE(\boldx)$ is a codeword, but different from
$\boldE(\boldx)$. We show now that this event can be excluded by
choosing $\epsilon$ small enough. For any two words ${\mathbf
F},{\mathbf G}$ we have $m-\langle {\mathbf F}, {\mathbf G}\rangle
\geq d({\mathbf F},{\mathbf G})$. If ${\mathbf F},{\mathbf G}$ are
codewords we also have $d({\mathbf F},{\mathbf
G})\geq(1-\Delta)m$, which implies then that $\langle {\mathbf F},
{\mathbf G}\rangle\leq m\Delta$. As we have seen earlier,
$\langle\boldE(\boldx),\tilde\boldE(\boldx)\rangle\geq m
\sqrt{1-\epsilon}$. Thus, choosing $\epsilon \leq 1-\Delta^2$
ensures that $\tilde\boldE(\boldx)$ cannot be a codeword.
\end{proof}

This property has two main consequences. The first is that the
existence of a complex state of the form $|h_\boldx\rangle$
follows immediately from the existence of a complex string. The
second is the fact that there exists a ``recipe'' that allows us
to prepare complex states, or would allow us to do so if we could
have a classical complex string.

\begin{cor}
For any $m\geq 1$ there exists a $(\log m +1)$-qubit complex state
of the form $|h_\boldx\rangle=\sum_{i=1}^m
|i\rangle|E_i(\boldx)\rangle$, where $\langle
i|j\rangle=\delta_{ij}$,
$x_i\in\{0,1\}$, and $E$ is an error-correcting code of the type introduced above.\\
\end{cor}

\section{Applications for quantum Kolmogorov complexity}
In the following we apply the results on quantum Kolmogorov
complexity to prove some statements from the theory of communication and
computation complexity. While some of the results are known, we underline that the proofs are simpler than those found
in literature. Finally we see how the definition of network complexity can be applied to some aspects in thermodynamics, allowing us to generalize existing results.

\subsection{Communication complexity}

Let us first show that the exponential upper bound on the
Kolmogorov complexity of quantum states implies that any quantum
communication protocol in the SMP model, requiring $q$ qubits of
communication, can be simulated using at most $\mco(2^q)$
classical bits of communication. Such result was already known
\cite{shizhu}, but here we can clearly see how it follows
 from basic considerations related to the known
properties of Kolmogorov complexity.

\begin{prop}
For any function $f:\{0,1\}^n\to\{0,1\}$, the gain in
communication complexity that one can obtain by using quantum
instead of classical communication, if only one-way communication
is allowed, is at most exponential, i.e. \beq \label{upperbound}
C_\cl(f)\leq \mco(2^{C_Q(f)}). \eeq
\end{prop}
\begin{proof}
The idea of the proof is to simulate the optimal quantum protocol
classically. Consider an optimal quantum communication protocol to
evaluate $f(x,y)$, and suppose that a single state of $q$ qubits
is sent. Such a state can be described by a classical sequence of
at most $\mco(2^q)$ bits, and any operation on the quantum state
can be simulated. In this case, thus,
$C_\cl(f)\leq\mco(2^{C_Q(f)})$. If instead of a single $q$-qubit
state, k states of $q'$ qubits are sent (with $kq'=q$), they can
be described by classical sequences of at most $\mco(2^{q'})$
qubits each, thus $C_\cl(f)\leq \mco(k 2^{q'})\leq
\mco(2^q)=\mco(2^{C_Q(f)})$.
\end{proof}

Note that a similar argument can be used to consider the more general case where also two-way communication between Alice and Bob is allowed.
Similarly as before, one only has to consider the classical
simulation of a quantum protocol. However, here it could be that
Alice does not send all the qubits to Bob which she uses to solve
the problem.

An exponential quantum/classical gap in a quantum communication
problem, is obtained only when in Eq. (\ref{upperbound}) we have
an equality (an exponential gap, in fact, requires
$C_Q(f)\leq\mco(\log C_\cl(f))$). This can be obtained only when
at least one of the states that appear in the quantum protocol is
complex (that is, its classical description grows exponentially
with the number of qubits). This implies that the power of quantum
communication can be explained (mainly) by quantum entanglement.
To see this, let us use for instance the network complexity
(Eq.(\ref{DefNetwork})). In \cite{prlnostro} it has been proved
that the entanglement of a quantum state, in terms of its Schmidt measure, is an upper bound for its
complexity. This implies that in all communication complexity
problems where one sees an exponential classical/quantum gap,
maximally entangled states must be created. This seems to imply
that the source of the quantum advantage is in fact entanglement.

\subsection{Computation complexity}

Quantum Kolmogorov complexity can also be used as a powerful tool
to gain insight on some issues related to computation complexity
theory. We use it here to determine some properties of the states
that appear during the computation.

Let us describe a quantum algorithm by a sequence of single- and
two-qubit operations, $A_{i_t,j_t}$, where $i_t$ and $j_t$ denote
the two qubits on which the operation is acting during the $t$-th time step (if $i_t=j_t$
$A_{i_t,j_t}$ denotes a single qubit operation acting on qubit
$i_t$). In the $t$-th step of the algorithm, the state
$|\Phi_{t-1}\rangle$ is transformed to
$|\Phi_t\rangle=A_{i_t,j_t}|\Phi_{t-1}\rangle$, with
$\ket{\Phi_0}=\ket{0}^{\otimes n}$. We have the following

\begin{obs}
If ${\mathcal{A}}=(A_{i_1,j_1},A_{i_2,j_2},\cdots,A_{i_T,j_T})$ is
a quantum algorithm with running time $T=\mco(\textrm{poly}(n))$,
then all the states $|\Phi_t\rangle=A_{i_t,j_t}|\Phi_{t-1}\rangle$
must have network Kolmogorov complexity that grows at most polynomially with
$n$.
\end{obs}
\begin{proof}
\noindent Since the quantum circuit model is universal, any
algorithm with running time $T$ can be simulated efficiently by a
circuit. In particular this implies that, for each step $t$, there
exists a quantum circuit that prepares $|\Phi_t\rangle$ requiring
at most $\mco(\textrm{poly}(t))$ gates. As the network complexity
of a state is trivially bounded from above by the size of the circuit, that is, the number of gates required
to
prepare it, and $t\leq T =\mco(\textrm{poly}(n))$, it follows that, for all $t$, $K_{\net}^\epsilon(|\Phi_t\rangle)\leq\mco(\textrm{poly}(n))$.\\
\end{proof}

On the other hand, one can show that if the complexities of the
states occurring in the algorithm are independent of $n$, then the
quantum algorithm cannot have an exponential speed--up compared to
a classical algorithm. Similarly to the classical case, we define
$K^\epsilon_Q(\ket{\Psi}|n)$ to be the number of bits required to
prepare the $n$--qubit state $\ket{\Psi}$ given the number $n$.

\begin{prop}
Let ${\mathcal{A}}=(A_{i_1,j_1},A_{i_2,j_2},\cdots,A_{i_T,j_T})$
be a quantum algorithm with running time
$T=\mco(\textrm{poly}(n))$, and let
$|\Phi_t\rangle=A_{i_t,j_t}|\Phi_{t-1}\rangle$ denote the state at
the $t$-th step of the computation. If for all $t$
$K_{\net}^\epsilon(|\Phi_t\rangle|n)\leq r_t$, where $r_t$ is
independent of $n$, then ${\mathcal{A}}$ can be simulated
classically with only a polynomial overhead of operations.
\end{prop}
\begin{proof}
\noindent If $K_{\net}^\epsilon(|\Phi_t\rangle|n)\leq r_t$ then
$\ket{\Phi_t}=M_t\ket{0}^{\otimes n}$, where the dimension of the operator
$M_t$ is some function of $r_t$, $f_t(r_t)$, but independent of $n$.
Thus, the number of operations required to compute $\ket{\Phi_t}$
classically is independent of $n$. Therefore, the classical
algorithm: compute for each $t$ the state $\ket{\Phi_t}$ and print
it solves the same problem as the quantum algorithm and requires
only $R T=R \mco(\textrm{poly}(n))$ computational steps, where $R$
is the maximum of all $f_t(r_t)$. \end{proof}

In \cite{JozsaLinden} it was proved that in order for a quantum
algorithm to be exponentially faster than a classical one that
solves the same task, it is necessary that at least one of the
states $\ket{\Phi_t}$ must describe an entangled state of $l$
qubits, where $l$ grows with the input size $n$. Together with the
results above, this means that, in order to find algorithms that
actually allow an exponential quantum/classical gap, we must look
for states that are both (a) highly entangled and (b)
have a non-constant Kolmogorov complexity that grows at most
polynomially with the input size. We underline the relevance of
this consideration which implies that, while entanglement is
undoubtfully a fundamental resource, it is not sufficient: only
particular types of entanglement are, in fact, useful.

\subsection{Kolmogorov complexity and thermodynamics}

As a last application of the notion of quantum Kolmogorov
complexity investigated in this paper let us reconsider some aspects
of thermodynamics \footnote{For a comprehensive analysis of the problem, and issues related to it, we refer to \cite{MaxwellDemon} and references therein.}. First of all we recall how the Kolmogorov
complexity became important in the context of thermodynamics and review some results presented by Caves \cite{Caves1,Caves2}
in this context. Then we show how these results can be
further developed by taking the notion of quantum Kolmogorov complexity into
account.

The second law of thermodynamics states that the entropy of a closed system, defined
as the logarithm of the number of microstates populated by a system,
 can never decrease. In his famous argument, Maxwell introduced a
ficticious machine where this law seemed to be violated. This machine
consists of a gas cylinder initially at equilibrium, and a
hypothetical demon which could open and close a shutter between  two parts 
of the cylinder. If the demon was able to separate the the fast molecules from the slow
ones, the entropy of the system could decrease, thus violating the second law of thermodynamics. 
The resolution to this paradox lies in the fact that the
demon must measure the velocity of the molecules and, more
importantly, that he must store this information in some memory. As
this memory has to be finite, the demon must erase the measurement
outcomes at some point in order to have space to store the new ones. It
could be shown from a physical point of view that this erasure
increases the entropy of the total system, ensuring that the
second law of thermodynamics is valid. Landauer formalized this fact
 in his famous principle, which states that
the process of erasing $n$ bits of information from a register
increases the entropy of the environment by $\log_2 n$
\cite{landauer}. Thus, sequences with large Kolmogorov complexity
cannot be erased, except with an irreversible process, which can
provide the required energy. Somehow inverse to the process of
erasure is the process of \emph{randomization}. Whenever a
computer makes a randomizing step (such as, e.g., tossing a coin)
it is possible to increase the information contained in the
computer register and thus, if used appropriately, a similar
procedure can be used to lower the entropy of the environment.

Bennett \cite{BennettThermo} and later  Zurek \cite{ZurekPRA} showed that
the Komogorov complexity of the microstates of a system is a
fundamental quantity in the thermodynamics analysis of such
problems. If the demon finds out a way to compress the bit string
he wants to erase, he might use a much lower thermodynamic cost
for the erasure. This deep connection between thermodynamic
entropy and Kolmogorov complexity has been formalized by Zurek in
\cite{ZurekPRA}, who has defined the \emph{total entropy} $\ovmcs$
of a system as the sum of the thermodynamic entropy and the
Kolmogorov complexity.

Caves continued the investigation of the Maxwell`s demon in the
quantum setting \cite{Caves1, Caves2}. He studied the change of
the total entropy, defined by Zurek as $\ovmcs=S+I$, where $S$
denotes the statistical entropy and $I$ the algorithmic
information. Here, the algorithmic information is composed of two
parts, the background information, which is sufficient to generate
a list of all states (up to a certain precision), and the
additional information, needed to describe a particular element of
this list of states (see Section II). In his considerations, Caves
chooses the background information to be negligible, implying that
only the conditional algorithmic information contributes to the
total entropy.

In order to demonstrate that it is indeed possible to conceive situations in which
an intelligent demon could extract work from a system (i.e. increase
the system's total entropy) Caves considered the following
situation. A demon, whose system is prepared in some standard product
state denoted by $\ket{\mathbf{s}}=\ket{s,\ldots,s}$, is acting on
a single photon, which is initially prepared in equilibrium,
$\rho_s = \one/2$. The total entropy of the initial state is
therefore
$\ovmcs_{\textrm{in}}=\ovmcs(\rho_s)+\ovmcs(\ket{\mathbf{s}}\bra{\mathbf{s}})=S(\rho_s)+I((\ket{\mathbf{s}}\bra{\mathbf{s}})=1$
\footnote{Here it is assumed, as mentioned above, that the
algorithmic information of the equilibrium state is $0$.}. The
main idea to increase the total entropy of the closed system is to
project the state of the photon onto an algorithmically complex
state. To this aim the demon randomizes his register (apart from
one bit) to obtain a complex string, $r$, of $m$ bits. He
associates to the bit string $r$ the angle $\theta$, where $0.0r$
is the binary representation (first $m$ bits) of $\theta/\pi$. The
precision with which he can choose one of the $2^m$ angles,
$\{\theta_k=k \frac{\pi}{2^m}\}_{k=0}^{2^m-1}$  is therefore
$\epsilon_\theta=2^{-m}\pi$. For any choice of $k$ (that is, $r$) he sets a
polarizing beam splitter to measure the polarization of the photon
in the basis $\{
\ket{\Psi_{\theta_k}}\equiv\cos(\theta_k)\ket{0}+\sin{\theta_k}\ket{1},
|\hat{\Psi}_{\theta_k}\rangle\equiv\sin(\theta_k)\ket{0}-\cos{\theta_k}\ket{1}
\}$. Thus, the state describing the system and the demon is $\rho=\sum_k
(\rho_{\theta_k}\otimes \proj{0}_{s}\otimes
\proj{\Psi_{\theta_k}}+\hat{\rho}_{\theta_k}\otimes
\proj{1}_{s}\otimes |\hat{\Psi}_{\theta_k}\rangle\langle\hat{\Psi}_{\theta_k}|) \otimes
\proj{e_k}_{d}$, where we have introduced an auxiliary system, $s$
that keeps track of information about which arm the photon left
the beam splitter and is available to the system. The auxiliary
system $d$ keeps track of all the possible measurement choices.
Here, the states $\ket{e_k}$ denote a orthonormal basis, and
$\rho_{\theta_k}=
\langle\Psi_{\theta_k}|\rho_S|\Psi_{\theta_k}\rangle$ and
$\hat{\rho}_{\theta_k}=
\langle\hat{\Psi}_{\theta_k}|\rho_S|\hat{\Psi}_{\theta_k}\rangle$
are not normalized. Thus, the final state describing the photon is
given by the reduced state of $\rho$, i.e. by tracing over system
$D,d$. Observing the measurement outcome the demon stores this
additional bit of information and thus holds the bit string $0.0r$
(or $0.1r$) respectively. From the demons point of view, the
system is after the measurement in a pure state,
$|\Psi_{\theta_k}\rangle$, or $|\hat{\Psi}_{\theta_k}\rangle$. The
total entropy of the whole system (w.l.o.g. we assume that $0.0r$ has
been measured) is therefore
$\ovmcs_{\textrm{fin}}=S(|\Psi_{\theta_k}\rangle)+I(\theta_k=0.0r)=I((0.0r))=m+1$.
Therefore, the total entropy is increased by $\Delta\ovmcs=m$,
which implies that the work $W^{(+)}=m k_BT\ln2$ has been
extracted.

An important point in these considerations is the presence of the
\emph{background information}, which is the information required
to fully describe the system \cite{CavesSchack}. In the example considered above, the
background information is the list of states ${\cal
S}\equiv\{\cos\theta_k|0\rangle+\sin\theta_k|1\rangle \mid
\theta_k=k \frac{\pi}{2^m} \}_{k=0}^{2^m-1}$ together with the uniform
probability distribution \footnote{Equivalently, one might
consider as background information the two basis states
$\{|0\rangle,|1\rangle\}$ and the set of angles $\{\theta_k=k
\frac{\pi}{m}\}_k$, where one agrees that the specification of an angle
$\theta_k$ implies preparing a state of the form
$\cos\theta_k|0\rangle+\sin\theta_k|1\rangle$.}.

In the following we shall choose a different point of view, for
two main reasons. First, while it is clear that a state can always
be described with the background information and the conditional
information, this might not always be the most convenient description.
Second, we see that with this approach it is very hard to go from
one system to composite systems. Consider the example studied above for
the case where the demon tries to measure $n$ photons. A possible
generalization of the background information in the case of $n$
systems is the set ${\cal S}^{\otimes n}$. In this case one would
still be allowed to consider the background information
negligible. At the same time, though, the demon does not have the
possibility to choose an arbitrary measurement direction with
arbitrary good precision, but is instead forced to project the
global system onto product states. The only situation in which the
demon could measure along arbitrary directions would be if the set
of its available states increased with the dimension of the
system. In fact, as seen before, only if his background
information increased exponentially with the number of qubits  the
demon could indeed have the freedom to measure an arbitrary state.
In this case it seems to be impossible to neglect the contribution
due to the background information.

If the demon can only measure each single photon independently
from all the others, the increase in total entropy is simply $n$
times the one that we have in the case of the single photon, i.e.
$\Delta\ovmcs = nm$. 
However, if the demon projects the system onto maximally complex
(and thus necessarily fully entangled) states the increase of the
total entropy can be \emph{exponentially larger}. The maximum
possible increase in the total entropy is given by
$\Delta\ovmcs^{(\textrm{Max})}=I_{\textrm{fin}}-NH_{\textrm{in}}=2^n\log\frac{1}{\epsilon}-n$,
where $\epsilon$ gives the precision with which the demon prepares
its states.

Note that for a general definition of the quantum Kolmogorov complexity
the background information is implicitely accounted for. This is
why then it is no problem to consider the projection onto complex
states. For instance using the network definition, the
background information is a basis consisting of single-qubit and
two--qubit gates, together with an initial tensor-product state.
In this case the background information remains the same, and
therefore negligible, when increasing the number of considered
qubits. In the case of the CBE complexity, instead, the background
information consists with the specification of the computational
basis: which is indeed algorithmically simple. This is why e.g.
the network, or the CBE, complexity of a state $\ket{\Psi}$ is
indeed the number of bits required to describe the state and no
additional information is required.


\section{Conclusions}
In this paper we investigated the generalization of the concept of
Kolmogorov complexity to the quantum case. We have shown how this
quantity can be related to other fundamental concepts of quantum
information theory, such as communication and computation
complexity. More precisely by analyzing a particular protocol of
communication complexity we found a condition that allows us to
discriminate among the different proposals for quantum Kolmogorov
complexity. In particular we have proven that any definition of
quantum Kolmogorov complexity that measures, in bits, the amount
of information needed to prepare a $n$--qubit state must scale
exponentially with $n$. We have also shown how some of the
existing definitions indeed satisfy such a condition. Furthermore,
it has been possible to use properties of the quantum Kolmogorov
complexity to prove statements about communication and computation
complexity. While some of the results we have obtained were
already known, the proofs based on quantum Kolmogorov complexity are
extremely simple, and encourage us to look for other possible
applications to related fields (such as query complexity
and oracle models for computation). Referring to the concept of
total entropy, first introduced by Zurek \cite{ZurekNature}, and
in particular to works by Caves \cite{Caves1, Caves2}, we
continued the study of the relation between quantum complexity and
thermodynamics. In particular, by applying the concept of quantum
Kolmogorov complexity, we were able to extend their analysis to a
full quantum setting, including a first study of the effects of
entanglement.

\section{Acknowledgements}
We would like to thank C. M. Caves for useful discussions.
We acknowledge support by the Austrian Science Foundation (FWF), the EU (OLAQUI, SCALA, QICS), and the Elise Richter Program.

\end{document}